\documentclass[a4paper,twoside,reqno,12pt]{article}
\usepackage{amsmath,amssymb}
\usepackage{amsthm}
 \newtheorem{thm}{Theorem}

 \newtheorem{prop}[thm]{Proposition}
  \newtheorem{defn}[thm]{Definition}
 \theoremstyle{remark}
 
 \newtheorem{rl}[thm]{Rule}
  \numberwithin{equation}{section}
\numberwithin{thm}{section}
\newtheorem{ex}[thm]{Example}

\begin{document}

\title{RECTANGLE GELL-MANN MATRICES} \author{} \maketitle

Christian RAKOTONIRINA 

\textit{Institut Sup\'{e}rieur de Technologie d'Antananarivo, IST-T,
BP 8122,}

Madagascar

E-mail: rakotopierre@refer.mg

\begin{abstract}
  We call rectangles Gell-Mann matrices rectangle matrices which make
  generalization of the expression of a tensor commutation matrix $n \otimes
  n$ in terms of tensor products of square Gell-Mann matrices.
\end{abstract}

\emph{\textbf{MCS 2000}}: \emph{Primary 15A69; secondary 15A90}

\section{Introduction}

\noindent For any $n \in \mathbb{N}$, $n \geqslant 2$, a tensor
commutation matrix $n \otimes n$, can be expressed in terms of $n
\times n$-Gell-Mann matrices $\Lambda_a^{( n )}$ for the following
way {\cite{rak051}}
\begin{equation}
  \label{eq1} U_{n \otimes n} = \frac{1}{n} I_n \otimes I_n + \frac{1}{2}
  \sum_{a = 1}^{n^2 - 1} \Lambda_a^{( n )} \otimes \Lambda_a^{( n )}
\end{equation}
where $I_{n}$ is the $n \times n$ unit matrix. Particularly, for $n
= 2$, $U_{2 \otimes 2}$ can be expressed in terms of the Pauli
matrices, namely
\[ U_{2 \otimes 2}=\left( \begin{array}{cccc}
     1 & 0 & 0 & 0\\
     0 & 0 & 1 & 0\\
     0 & 1 & 0 & 0\\
     0 & 0 & 0 & 1
   \end{array} \right) = \frac{1}{2} I_2 \otimes I_{2} + \frac{1}{2} \sum_{i
   = 1}^3 \sigma_i \otimes \sigma_i \]
which is frequently found in quantum information theory
{\cite{fuj01}}, {\cite{fad95}}, {\cite{fra02}}.

\noindent We had tried {\cite{rak051}} to express the tensor
commutation matrix $U_{2 \otimes 3}$ and $U_{3 \otimes 2}$ as linear
combinations of the tensor products of the Pauli matrices with the
$3 \times 3$-Gell-Mann matrices, in expecting to have expressions
that lead to the generalization of (\ref{eq1}). However, the
obtained expressions are not interesting enough for the wanted
generalization. We have noticed that for generalizing (\ref{eq1}) to
the expression of $U_{n \otimes p}$, $n \neq p$, we should use
rectangle matrices instead of square matrices. We call rectangle
Gell-Mann matrices such rectangle matrices.

At first, we talk about the tensor commutation matrices. After that,
we construct the $n \times p$-Gell-Mann matrices, in starting with
some examples.

Let us denote by $I_{n \times p}$ the $n \times p$ matrices obtained
in adding into the unit matrix $I_{\inf( n, p )} $ $|n - p|$ rows or
$|n - p|$ columns formed by zeros.

\section{Tensor commutation matrix}

\begin{defn}
  For $n, p \in \mathbb{N}$, $n \geqslant 2$, $p \geqslant 2$, we call tensor
  commutation matrix $n \otimes p$ the permutation matrix $U_{n \otimes p}\in
  \mathcal{M}_{np \times np}(\mathbb{C})$ formed by $0$ and
  $1$, verifying the relation
  \[ U_{n \otimes p} \cdot ( a \otimes b ) = b \otimes a \]
  for all $a \in \mathcal{M}_{n \times 1} (\mathbb{C})$, $b \in
  \mathcal{M}_{p \times 1} (\mathbb{C})$.
\end{defn}

We can construct $U_{n \otimes p}$ by using the following rule
{\cite{rak050}}.

\begin{rl}

\textit{ {\noindent}Let us start in putting $1$, at the first row
and first column, after that let us pass into the second column in
going down at the rate of $n$ rows and put $1$ at this place, then
pass into the third column in going down at the rate $n$ rows and
put $1$, and so on until there is only $n - 1$ rows for going down (
then we have obtained as numbere of $1$: $p$ ). Then, pass into the
next column which is the $( p + 1 )$-th column, put $1$ at the
second row of this column ( second row, because $( n - 1 ) + 1 = n
)$ and repeat the process until we have $n - 2$ rows for going down
( then we have obtained as numbere of $1$: $2 p$ ). After that, pass
into the next column which is the$( 2 p + 1 )$-th column, put $1$ at
the third row of this column ( third row, because $( n - 2 ) + 2 =
n$ ) and repeat the process until we have $n - 3$ rows for going
down ( then we have obtained as number of $1$: $3 p$ ). Continuing
in this way we will have that the element at $n \times p$-th row and
$n \times p$-th column is $1$. The other elements are $0$. }
\end{rl}

\begin{prop}
  For $n, p \in \mathbb{N}$, $n, p \geqslant 2$,
  \[ U_{n \otimes p} = \sum_{( i, j )}^{( p, n )} E_{p \times n}^{( i, j )}
     \otimes E_{p \times n}^{( i, j )^t} = \sum_{( i, j )}^{( p, n )} E_{p
     \times n}^{( i, j )} \otimes E_{n \times p}^{( j, i )} \]
  where $E_{p \times n}^{( i, j )}$ is the elementary $p \times n$-matrix
  formed by zero except the element at $i$-th row, $j$-th column which
  is equal $1$.
  \end{prop}

\begin{proof}
  Let $a = \left( \begin{array}{c}
    a_1\\
    a_2\\
    \vdots\\
    a_n
  \end{array} \right) \in \mathcal{M}_{n \times 1}(\mathbb{C})$, $b =
  \text{$\left( \begin{array}{c}
    b_1\\
    b_2\\
    \vdots\\
    b_p
  \end{array} \right) \in \mathcal{M}_{p \times 1}(\mathbb{C})$}$
\begin{equation*}
\begin{split}
   U_{n \otimes p} \cdot(a \otimes b) &= \sum_{( i, j )}^{( p, n )}
     E_{p \times n}^{( i, j )} \otimes E_{n \times p}^{( j, i )} \cdot ( a
     \otimes b )\\
      &\quad= \sum_{( i, j )}^{( p, n )} ( E_{p \times n}^{( i, j )}
     \cdot a ) \otimes ( E_{n \times p}^{( j, i )} \cdot b )\\
     &\quad= \sum_{( i, j )}^{( p, n )} ( \delta_{ik} a_j )_{1 \leqslant k \leqslant
     p} \otimes ( \delta_{jl} b_i )_{1 \leqslant l \leqslant n}\\
     &\quad= \sum_{( i, j )}^{( p, n )} ( \delta_{ik} b_i )_{1 \leqslant k \leqslant
     p} \otimes ( \delta_{jl} a_j )_{1 \leqslant l \leqslant n}\\
     &\quad= \sum_{( i, j )}^{( p, n )} \left( \begin{array}{c}
       0\\
       \vdots\\
       0\\
       b_i\\
       0\\
       \vdots\\
       0
     \end{array} \right) \otimes \left( \begin{array}{c}
       0\\
       \vdots\\
       0\\
       a_j\\
       0\\
       \vdots\\
       0
     \end{array} \right)\\
     &\quad= b \otimes a
\end{split}
\end{equation*}
  \begin{ex}
    The application of the rule yields us

\begin{equation*}
       U_{2 \otimes 3} = \left( \begin{array}{ccccccc}
         1 & 0 & 0 & 0 & 0 & 0 & \\
         0 & 0 & 0 & 1 & 0 & 0 & \\
         0 & 1 & 0 & 0 & 0 & 0 & \\
         0 & 0 & 0 & 0 & 1 & 0 & \\
         0 & 0 & 1 & 0 & 0 & 0 & \\
         0 & 0 & 0 & 0 & 0 & 1 &
       \end{array} \right)
       \end{equation*}
       \begin{equation*}
       \begin{split}
    U_{2 \otimes 3} &= \left( \begin{array}{cc}
      1 & 0\\
      0 & 0\\
      0 & 0
    \end{array} \right) \otimes \left( \begin{array}{ccc}
      1 & 0 & 0\\
      0 & 0 & 0
    \end{array} \right) + \left( \begin{array}{cc}
      0 & 1\\
      0 & 0\\
      0 & 0
    \end{array} \right) \otimes \left( \begin{array}{ccc}
      0 & 0 & 0\\
      1 & 0 & 0
    \end{array} \right) + \left( \begin{array}{cc}
      0 & 0\\
      1 & 0\\
      0 & 0
    \end{array} \right) \otimes \left( \begin{array}{ccc}
      0 & 1 & 0\\
      0 & 0 & 0
    \end{array} \right)\\
     &+ \left( \begin{array}{cc}
      0 & 0\\
      0 & 1\\
      0 & 0
    \end{array} \right) \otimes \left( \begin{array}{ccc}
      0 & 0 & 0\\
      0 & 1 & 0
    \end{array} \right) + \left( \begin{array}{cc}
      0 & 0\\
      0 & 0\\
      1 & 0
    \end{array} \right) \otimes \left( \begin{array}{ccc}
      0 & 0 & 1\\
      0 & 0 & 0
    \end{array} \right) + \left( \begin{array}{cc}
      0 & 0\\
      0 & 0\\
      0 & 1
    \end{array} \right) \otimes \left( \begin{array}{ccc}
      0 & 0 & 0\\
      0 & 0 & 1
    \end{array} \right)
    \end{split}
\end{equation*}
  \end{ex}
\end{proof}

\section{Construction of a system of rectangle Gell-Mann matrices}

At first, let us consider some particular cases.

\subsection{$U_{2 \otimes 3}$}

The $2 \times 2$-Gell-Mann matrices are the Pauli matrices $\sigma_1
= \left(
\begin{array}{cc}
  0 & 1\\
  1 & 0
\end{array} \right)$,\\
 $\sigma_{2} = \left( \begin{array}{cc}
  0 & - i\\
  i & 0
\end{array} \right)$, $\sigma_{3} = \left( \begin{array}{cc}
  1 & 0\\
  0 & - 1
\end{array} \right)$.

Let $\sigma_0 = \left( \begin{array}{cc}
  1 & 0\\
  0 & 1
\end{array} \right)$. Being inspired by a way for constructing the $n \times
n$-Gell-Mann matrices from the $( n - 1 ) \times ( n - 1
)$-Gell-Mann matrices, where on the diagonal there is not zero
between two non zero elements and the first non zero element is the
first element ( that is, the element at the first row, first column
) (Cf. for example{\cite{rak060}}), we add into these four matrices
third column formed by zeros. Then, we have a system formed by
\begin{equation*}
I_{2 \times 3} = \left( \begin{array}{ccc}
  1 & 0 & 0\\
  0 & 1 & 0
\end{array} \right), \Lambda_1 = \left( \begin{array}{ccc}
  0 & 1 & 0\\
  1 & 0 & 0
\end{array} \right), \Lambda_2 = \left( \begin{array}{ccc}
  0 & - i & 0\\
  i & 0 & 0
\end{array} \right),
 \Lambda_3 = \left( \begin{array}{ccc}
  1 & 0 & 0\\
  0 & - 1 & 0
\end{array} \right)
\end{equation*}
 And for obtaining a basis of $\mathcal{M}_{2 \times 3}(
\mathbb{C})$, we introduce the matrices\\
 $\Lambda_4 = \left(
\begin{array}{ccc}
  0 & 0 & \sqrt{2}\\
  0 & 0 & 0
\end{array} \right), \Lambda_5 = \left( \begin{array}{ccc}
  0 & 0 & 0\\
  0 & 0 & \sqrt{2}
\end{array} \right)$. We can check easily that
\[ U_{2 \otimes 3} = \frac{1}{2} I_{2 \times 3}^+ \otimes I_{2 \times 3} +
   \frac{1}{2} \sum_{a = 1}^5 \Lambda_a^+ \otimes \Lambda_a \]
where $\Lambda_a^+$ is the hermitian conjugate of $\Lambda_a$.

\subsection{$U_{3 \otimes 2}$}

Using analogous way, but this time we are adding into the Pauli
matrices and $\sigma_0$ a third row formed by zeros, instead of
column. Then, we get a system $( \Lambda_a )_{1 \leqslant a
\leqslant 5}$ of $3 \times 2$ matrices which satisfies
\[ U_{3 \otimes 2} = \frac{1}{2} I_{3 \times 2}^+ \otimes I_{3 \times 2} +
   \frac{1}{2} \sum_{a = 1}^5 \Lambda_a^+ \otimes \Lambda_a \]
In fact, $U_{n \otimes p} = U_{n \otimes p}^t = U_{n \otimes p}^+$,
for all $n, p \in \mathbb{N}$, $n, p \geqslant 2$.

\subsection{$U_{2 \otimes 4}$}

Using yet the analogous way, but in this case we are adding into the
Pauli matrices and the $2 \times 2$ unit matrix third and fourth
columns formed by zeros. Then, we have a system formed by four $2
\times 4$ matrices $I_{2 \times 4}, \Lambda_1, \Lambda_2,
\Lambda_3$. And for obtaining a basis of $\mathcal{M}_{2 \times 4} (
\mathbb{C} )$, we introduce the matrices\\
\noindent $\Lambda_4 = \left(
\begin{array}{cccc}
  0 & 0 & \sqrt{2} & 0\\
  0 & 0 & 0 & 0
\end{array} \right)$, $\Lambda_5 = \left( \begin{array}{cccc}
  0 & 0 & 0 & 0\\
  0 & 0 & \sqrt{2} & 0
\end{array} \right)$, $\Lambda_6 = \left( \begin{array}{cccc}
  0 & 0 & 0 & \sqrt{2}\\
  0 & 0 & 0 & 0
\end{array} \right)$, $\Lambda_7 = \left( \begin{array}{cccc}
  0 & 0 & 0 & 0\\
  0 & 0 & 0 & \sqrt{2}
\end{array} \right)$ in the system. The system satisfies the relation
\[ U_{2 \otimes 4} = \frac{1}{2} I_{2 \times 4}^+ \otimes I_{2 \times 4} +
   \frac{1}{2} \sum_{a = 1}^7 \Lambda_a^+ \otimes \Lambda_a \]

\subsection{$U_{3 \otimes 4}$}

In this case, we start from the $3 \times 3$-Gell-Mann matrices.
Then, we have a system $( \Lambda_a )_{1 \leqslant a \leqslant 11}$
of $3 \times 4$ matrices which satisfies the relation
\[ U_{3 \otimes 4} = \frac{1}{3}_{} I_{3 \times 4}^+ \otimes I_{3 \times 4} +
   \frac{1}{2} \sum_{a = 1}^{11} \Lambda_a^+ \otimes \Lambda_a \]
\begin{defn}
  Let $n, p \in \mathbb{N}$, $p \geqslant n \geqslant 2$. We call $n \times
  p$-Gell-Mann matrices the $n \times p$ matrices $\Lambda_1$, $\Lambda_2$,
  ..., $\Lambda_{n^2 - 1}$, $\Lambda_{n^2}$, $\Lambda_{n^2 + 1}$, ...,
  $\Lambda_{np - 1}$ such that:

  $\Lambda_1$, $\Lambda_2$, ..., $\Lambda_{n^2 - 1}$ are obtained in adding
  into the $n \times n$-Gell-Mann matrices $( n + 1 )$-th, $( n + 2 )$-th,
  ..., $p$-th columns, formed by zeros;

  $\Lambda_{n^2} = \sqrt{2} E_{n \times p}^{( 1, n + 1 )}$, $\Lambda_{n^2 + 1}
  = \sqrt{2} E_{n \times p}^{( 2, n + 1 )},$..., $\Lambda_{n^2 + n - 1} =
  \sqrt{2} E_{n \times p}^{( n, n + 1 )}$,

  $\Lambda_{n^2 + n} = \sqrt{2} E_{n \times p}^{( 1, n + 2 )}$, $\Lambda_{n^2
  + n + 1} = \sqrt{2} E_{n \times p}^{( 2, n + 2 )}$, ..., $\Lambda_{n^2 + 2 n
  - 1} = \sqrt{2} E_{n \times p}^{( 1, n + 2 )}$,

  ....................................................................................................................,

  $\Lambda_{n^{} ( p - 1 )} = \sqrt{2} E_{n \times p}^{( 1, p )}$,
  $\Lambda_{n^{} ( p - 1 ) + 1} = \sqrt{2} E_{n \times p}^{( 2, p )}$, ...,
  $\Lambda_{n^{} p - 1} = \sqrt{2} E_{n \times p}^{( n, p )}$.

  Then, we define the $p \times n$-Gell-Mann matrices as the matrices obtained
  in taking the hermitian conjugates of the $n \times p$-Gell-Mann matrices.
\end{defn}

\begin{prop}
  For $n, p \in \mathbb{N}$, $p, n \geqslant 2$, consider the system of $n
  \times p$-Gell-Mann matrices $\Lambda_1$, $\Lambda_2$, ..., $\Lambda_{np -
  1}$. Then,
  \begin{equation}
    \label{eq2} U_{n \otimes p} = \frac{1}{\inf( n, p )} I_{n \times
    p}^+ \otimes I_{n \times p} + \frac{1}{2} \sum_{a = 1}^{np - 1}
    \Lambda_a^+ \otimes \Lambda_a
  \end{equation}
\end{prop}

\begin{proof}
  Let us suppose $p \geqslant n$.
  \begin{equation}
    \label{eq3} \frac{1}{2} \sum_{a = n^2}^{np - 1} \Lambda_a^+ \otimes
    \Lambda_a = \sum_{( j, l ) = ( 1, n + 1 )}^{( n, p )} E_{n \times p}^{( j,
    l )^t} \otimes E_{n \times p}^{( j, l )^{}}
  \end{equation}
  Using the proposition and the formula (\ref{eq1}) we have
  \begin{equation}
    \label{eq4} \sum_{( j, l ) = ( 1, 1 )}^{( n, n )} E_{n \times n}^{( j, l
    )^t} \otimes E_{n \times n}^{( j, l )^{}} = \frac{1}{n} I_n \otimes I_n +
    \frac{1}{2} \sum \Lambda_a^{( n )} \otimes \Lambda_a^{( n )}
  \end{equation}
  In adding, in (\ref{eq4}), into the terms on the left of $\otimes$'s $p - n$
  rows, $( n + 1 )$-th, $( n + 2 )$-th, ..., $p$-th rows, and on the right $p
  - n$ columns, $( n + 1 )$-th, $( n + 2 )$-th, ..., $p$-th columns, formed by
  zeros, the non zero elements which have same position, same row and same
  column, will keep same position. So by the definition of tensor product of
  matrices
  \[ \sum_{( j, l ) = ( 1, 1 )}^{( n, p )} E_{n \times p}^{( j, l )^t} \otimes
     E_{n \times p}^{( j, l )^{}} - \sum_{( j, l ) = ( 1, n + 1 )}^{( n, p )}
     E_{n \times p}^{( j, l )^t} \otimes E_{n \times p}^{( j, l )^{}} =
     \frac{1}{n} \Lambda_0^+ \otimes \Lambda_0 + \frac{1}{2} \sum_{a = 1}^{n^2
     - 1} \Lambda_a^+ \otimes \Lambda_a \]
  Using the proposition and (\ref{eq3}) we have (\ref{eq2}).
\end{proof}

Now, we are giving some properties of the rectangle Gell-Mann
matrices.

\begin{prop}
  For $n, p \in \mathbb{N}$, $p, n \geqslant 2$, let
$(\Lambda_a)_{1\leq a\leq np-1}$ a system of $n\times p$-Gell-Mann
matrices. Then,  $Tr (\Lambda_a^+ \Lambda_b) = 2 \delta_{ab}$
  where $\delta_{ab}$ is the Kronecker symbol.
\end{prop}

\begin{prop}
For $n, p \in \mathbb{N}$, $p\geqslant n \geqslant 2$, let
$(\Lambda_a)_{1\leq a\leq np-1}$ a system of $n\times p$-Gell-Mann
matrices. Then,
\begin{equation*}
     \Lambda_a^{} \Lambda^+_b - \Lambda_b^{} \Lambda^+_a = i\sum_{c=1}^{n^{2}-1}f_{abc} \Lambda^{(
     n )}_c
     \end{equation*}
     where $f_{abc}$'s the components of a tensor totally antisymmetric, with
     $f_{abc}=0$ if at least one of $a$, $b$, $c$ is in $\{n^{2}, n^{2}+1, ...,
     np-1\}$.
  \end{prop}

\section*{Conclusion}

Being inspired by a way for constructing the $n \times n$-Gell-Mann
matrices from the $( n - 1 ) \times ( n - 1 )$-Gell-Mann matrices,
where on the diagonal there is not zero between two non zero
elements and the first non zero element is the first element ( that
is, the element at the first row, first column ), we can construct a
basis of $\mathcal{M}_{n \times p} (\mathbb{C})$, whose  elements
make generalization of the expression of $U_{n \otimes n}$ in terms
of the tensor products of  $n \times n$-Gell-Mann matrices to the
expression of $U_{n \otimes p}$.


\begin{thebibliography}{9}
 \bibitem {rak051} RAKOTONIRINA.C, International Journal of Mathematics and Mathematical
 Sciences, Volume 2007, Article ID 20672, 10 pages, 2007.
  \bibitem{fuj01} FUJII.K, arXiv: quant-ph/0112090, prepared for 10th Numazu
  Meeting on Integral System, Noncommutative Geometry and
  Quantum theory, Numazu, Shizuoka, Japan, Mars 2002.
  \bibitem{fad95} FADDEV.L.D, Int.J.Mod.Phys.A, Vol.10, No 13, May,1848 (1995).
  \bibitem{fra02}VERSTRAETE.F, Th\`{e}se de Doctorat, Katholieke
  Universiteit Leuven, (2002).
  \bibitem{rak050} RAKOTONIRINA.C, arXiv: math.GM/0508053.
  \bibitem{rak060} RAKOTONIRINA.C, arXiv: hep-th/0601232.
  \end{thebibliography}
\end{document}